\documentclass[12pt]{iopart}
\usepackage{amssymb,epsfig,graphicx}
\begin{document}

\title[Absence of pre-transition phases in Ni$_{50}$Mn$_{25+x}$In$_{25-x}$]{Lattice strain accommodation and absence of pre-transition phases in Ni$_{50}$Mn$_{25+x}$In$_{25-x}$}

\author{R. Nevgi$^1$, K. R. Priolkar$^1$, L. Righi$^2$, M. Solzi$^3$, F. Cugini$^3$, E. Dias$^4$ and A. K. Nigam$^4$}
\address{$^1$Department of Physics, Goa University, Taleigao Plateau, Goa 403206 India}
\address{$^2$Department of Chemistry, Parma University, Parco Area delle Scienze 17/a 43124 Parma, Italy}
\address{$^3$Department of Mathematical, Physical and Computer Sciences, University of Parma, Parco Area delle Scienze 7/A 43124 Parma, Italy.}
\address{$^4$Department of Condensed Matter Physics and Material Science, Tata Institute of Fundamental Research, Colaba, Mumbai 400005 India}
\ead{krp@unigoa.ac.in}
\vspace{10pt}
\begin{indented}
\item[]July 2020
\end{indented}

\begin{abstract}
The stoichiometric Ni$_{50}$Mn$_{25}$In$_{25}$ Heusler alloy transforms from a stable ferromagnetic austenitic ground state to an incommensurate modulated martensitic ground state with a progressive replacement of In with Mn without any pre-transition phases. The absence of pre-transition phases like strain glass in Ni$_{50}$Mn$_{25+x}$In$_{25-x}$ alloys is explained to be the ability of the ferromagnetic cubic structure to accommodate the lattice strain caused by atomic size differences of In and  Mn atoms. Beyond the critical value of $x$ = 8.75, the alloys undergo martensitic transformation despite the formation of ferromagnetic and antiferromagnetic clusters and the appearance of a super spin glass state.
\end{abstract}

\maketitle

\section{Introduction}

A strain glass phase is expected to emerge in a ferroelastic/martensitic material with a dopant concentration greater than a critical value. The presence of impurities tends to curtail the long range order of the strain vector leading to a frozen disordered ferroelastic phase. This phase is akin to cluster glassy phase in impurity doped magnetic materials \cite{Wang201298,Sun2019183} or relaxor ferroelectrics \cite{Chen200291,Badapanda20143,Lu20177}. Several reports depicting the existence of strain glass phase have been reported in the literature \cite{Ren902010,Zhang812010}.  Nevertheless, several questions remain unanswered for example, \textit{do all martensitic alloys doped with defects display a strain glass transition?} Recently we have shown that in magnetic Heusler alloys undergoing martensitic transformation, site occupancy of the dopant plays a vital role in determining the alloy ground state \cite{Nevgi2018112,Nevgi2019797}. In Fe doped Ni-Mn-In alloys, when Mn is replaced by Fe, the suppression of the martensitic phase occurs via a strain glassy phase \cite{Nevgi2018112}. On the contrary, when Fe is substituted for Ni, the resulting ground state is a cubic ferromagnet \cite{Nevgi2019797}.

Ni$_{50}$Mn$_{50}$ is an antiferromagnet with tetragonal L1$_0$ structure in the martensite state below 973 K \cite{Kren196829}. A systematic replacement of Mn with In, Sn or Sb, results in a decrease of martensitic transition temperature $T_M$ eventually leading to complete suppression of the martensite phase above a critical concentration \cite{Sutou200485,Planes200921}. This critical concentration, as well as the rate of variation of transformation temperature with average electron per atom (e/a) ratio depends on the type of replaced atom (In, Sn or Sb). However, no non-ergodic phases like strain glass have been hitherto reported.

In literature, such Ni-Mn based magnetic shape memory alloys have been investigated as Ni$_{50}$Mn$_{25+x}$Z$_{25-x}$ (Z = In, Sn or Sb) as potential actuators due to their ability to exhibit large magnetic field induced strain in the martensitic state \cite{Chernenko199533, Ullakko199669}. A variety of other magneto-structural effects like magnetic superelasticity \cite{Sutou200485, Krenke200542, Krenke200673, Krenke200775, Kainuma2006957}, magnetocaloric effects \cite{Krenke20054, Moya200775}, giant magnetoresistance \cite{Chatterjee200942}, exchange bias \cite{Khan200791}, and kinetic arrest \cite{Ito200892, Sharma200776} have been reported paving the way for a new range of research possibilities.

In Ni$_{50}$Mn$_{25+x}$Z$_{25-x}$, with the lowering of temperature, the martensitic transformation is accompanied by magnetic transitions. While the high temperature austenite phase is predominantly ferromagnetic below the characteristic Curie temperature $T_C$, a state with competing ferro and antiferromagnetic interactions emerges in the martensitic state below the transformation temperature $T_M$ \cite{Aksoy200791,Siewert201214,Entel201365}. It is believed that the microscopic driving force for the martensitic transformation is the hybridization of Ni $d$ states with the antiferromagnetically coupled Mn $d$ states present in the Z atom site \cite{Ye2010104}. Diffraction studies have ascertained the martensitic structure to be the incommensurate 5M and 7M structures \cite{Righi200755, Righi200856, Yan201588}. In addition to the austenite-martensite transformation, intermartensitic transitions are observed on cooling depending on the alloy composition \cite{Asli201599}. EXAFS investigations have revealed the presence of local structural disorder that is believed to be responsible for the increased hybridization of the Ni $d$ and Mn $d$ states \cite{Bhobe200674, Bhobe200820, Lobo201096}. This strong coupling between the structural and magnetic degrees of freedom is a characteristic property witnessed in magnetostructural transition in Ni-Mn based Heusler alloys \cite{Orlandi942016}.

In such a backdrop of understanding the martensitic transition, a few questions remain unanswered. \textit{ Why does the martensitic transformation appear at a critical concentration of the dopant atom? Are there any pre-transition phases around this critical concentration that have hitherto not been discovered?} To answer these questions we have prepared the off stoichiometric compositions, Ni$_{50}$Mn$_{25+x}$In$_{25-x}$  near the critical concentration. In particular, we focus on Ni$_{50}$Mn$_{33}$In$_{17}$ ($x$ = 8), Ni$_{50}$Mn$_{33.75}$In$_{16.25}$ ($x$ = 8.75), Ni$_{50}$Mn$_{34.5}$In$_{15.5}$ ($x$ = 9.5) and Ni$_{50}$Mn$_{35}$In$_{15}$ ($x$ = 10). The stoichiometric Ni$_{50}$Mn$_{25}$In$_{25}$ does not undergo martensitic transition and exhibits a cubic ferromagnetic ground state. As the concentration of Mn increases, the martensitic transition appears at Ni$_{50}$Mn$_{33.75}$In$_{16.25}$ ($x$ = 8.75). By a careful study of the local structure and magnetic properties around the critical concentration, we investigate the cause for the appearance of martensitic transformation in Ni$_{50}$Mn$_{25+x}$In$_{25-x}$ without the presence of any non-ergodic pre-transition phases like the strain glass.

\section{Experimental}

The Ni$_{50}$Mn$_{25+x}$In$_{25-x}$ $(0 \leq x \leq 10)$ alloys were prepared by arc melting in an argon atmosphere of high purity elements (99.9\%). During the preparation process, the ingot of the individual alloy was flipped several times to ensure homogeneity. The ingots were then cut using a low speed diamond saw and powdered. The cut pieces and powders covered in tantalum foil were vacuum sealed in quartz tubes and annealed at 750$^\circ$C for 48 hours and subsequently ice quenched. The compositions were verified using scanning electron microscopy with energy dispersive x-ray (SEM-EDX) technique and were within 2\% of the targeted values. X-ray diffraction measurements were performed in the temperature range 80 K $< T <$ 300 K on an ARLX'TRA diffractometer in the 2$\theta$ range of 20$^\circ$ to 90$^\circ$ using Cu K$\alpha$ radiation. The data was analyzed by Rietveld method using Jana 2006 software \cite{Petricek2292014}. Temperature dependent magnetization measurements M(T) were carried out in the temperature range of 5 K -- 350 K. The samples were first cooled in zero applied magnetic field from room temperature to 5 K  and the data was recorded while warming (ZFC) followed by cooling (FCC) and subsequent warming (FCW). The isothermal magnetization measurements M Vs H were performed in the range $\pm$7 T. The samples were first cooled in zero field from 350 K to 5 K (ZFC-M(H)). Field cooled magnetization loops (FC-M(H)) were recorded by cooling the samples again from 350 K to 5 K in an applied field of 5T. AC magnetic susceptibility measurements were carried out in the temperature range 5 K -- 350 K at various excitation frequencies $(33 Hz \leq f \leq 9997 Hz)$ by applying AC magnetic field of  $H_{ac}$ = 10 Oe after cooling the sample in zero field in Physical Property Measurement Systems (Quantum Design, USA). The local structural studies were performed using Extended X-ray Absorption Fine structure (EXAFS)at Ni K (8333 eV) and Mn K (6539 eV) edges in the temperature range 50 K - 300 K at the P65 beamline (PETRA III Synchrotron Source, DESY, Hamburg, Germany). The incident ($I_0$) and the transmitted ($I$) photon energies were simultaneously recorded using gas ionization chambers as detectors. The thickness of the absorbers was adjusted by controlling the number of layers of scotch tape coated uniformly with alloy powders, to obtain the absorption edge jump $\Delta\mu(t) \leq 1$ where $\Delta\mu$ is the change in absorption coefficient at the absorption edge and $t$ is the thickness of the absorber. At each edge, at least three scans were collected to average statistical noise and analyzed using well established procedures in Demeter suite.\cite{Raval200512}

\section{Results}

\begin{figure}[h]
\begin{center}
\includegraphics[width=\columnwidth]{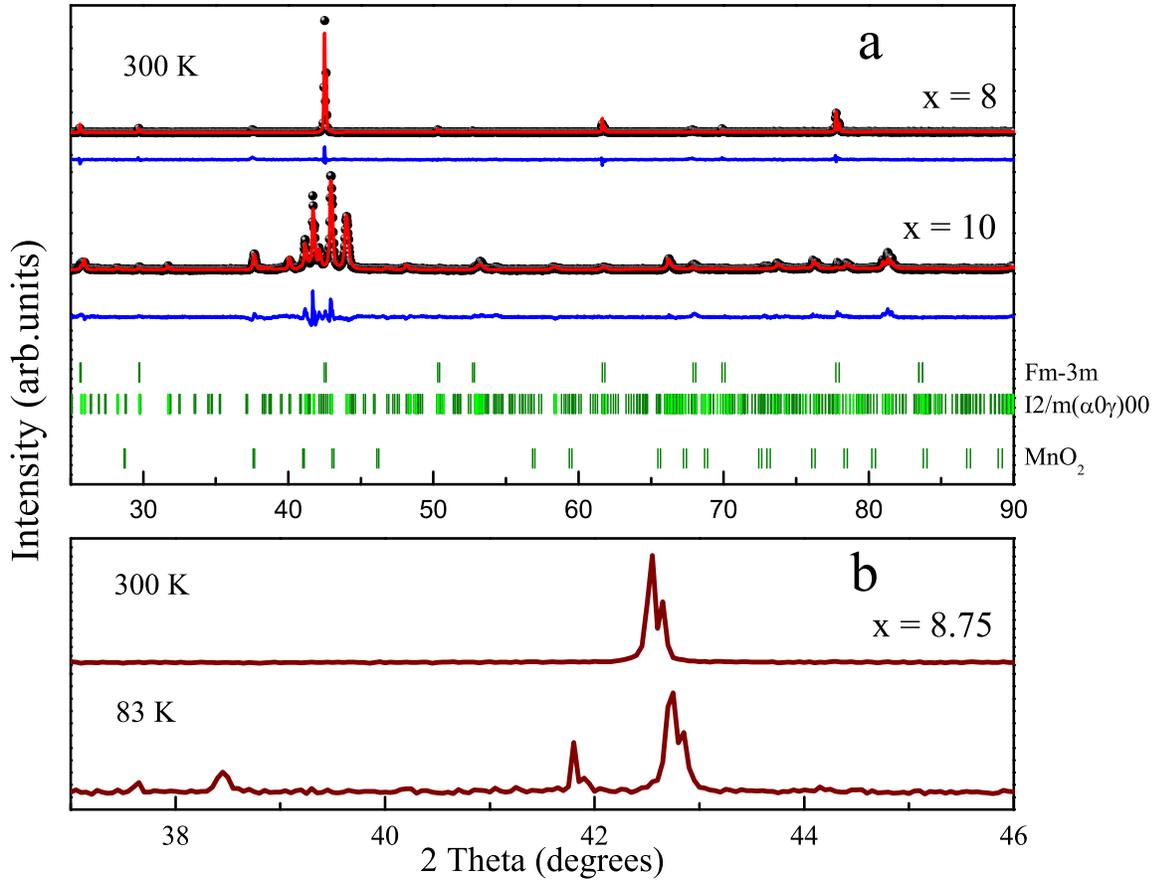}
\caption{Rietveld refined x-ray diffraction patterns of Ni$_{50}$Mn$_{25+x}$In$_{25-x}$, highlighting the cubic and modulated phases in the two alloys, $x$ = 8 and $x$ = 10 at 300K are shown in a. The transformation of $x$ = 8.75 alloy from austenitic structure at 300 K to one with coexisting martensitic and austenitic phases at 83 K is shown in the limited angular range ($37 \le 2\theta \le 46$) in b.}
\label{fig:XRD}
\end{center}
\end{figure}


The x-ray diffraction patterns for Ni$_{50}$Mn$_{25+x}$In$_{25-x}$ alloys are shown in Fig.\ref{fig:XRD}. The alloy, $x$ = 8 displays a cubic $L2_1$ structure ($Fm-3m$ space group) with lattice constant $a$ = 6.013 \AA~ at 300 K (Fig.\ref{fig:XRD} a) and retains its austenitic structure at all measured temperatures down to 80K. Fig.\ref{fig:XRD} b shows the x-ray diffraction data for $x$ = 8.75 alloy in the limited 2 $\theta$ range ($37 \le 2\theta \le 46$) displaying the 220 Bragg reflection of the cubic phase ($a$ = 6.007 \AA) at 300 K. Additional Bragg peaks appear in the diffraction pattern recorded below 150K indicating the martensitic transition. The martensitic phase coexists with the high temperature cubic austenitic phase down to the lowest temperature measured. On the other hand, the alloys $x$ = 9.5 and $x$ = 10 completely transform from their high temperature cubic $L2_1$ structure to 7M modulated martensitic structure as the temperature is lowered below their respective martensitic finish temperatures $M_F$. Fig.\ref{fig:XRD} a also shows the Rietveld refined data of the alloy $x$ = 10 at 300 K exhibiting 7M modulated structure solved using superspace approach with the space group $I2/m(\alpha0\gamma)00$ and lattice constants $a$ = 4.389 \AA, $b$ = 5.560 \AA, $c$ = 4.332 \AA and $\beta$ = 92.94$^{\circ}$ with a modulation vector $q$ = 0.338.

\begin{figure}[h]
\begin{center}
\includegraphics[width=\columnwidth]{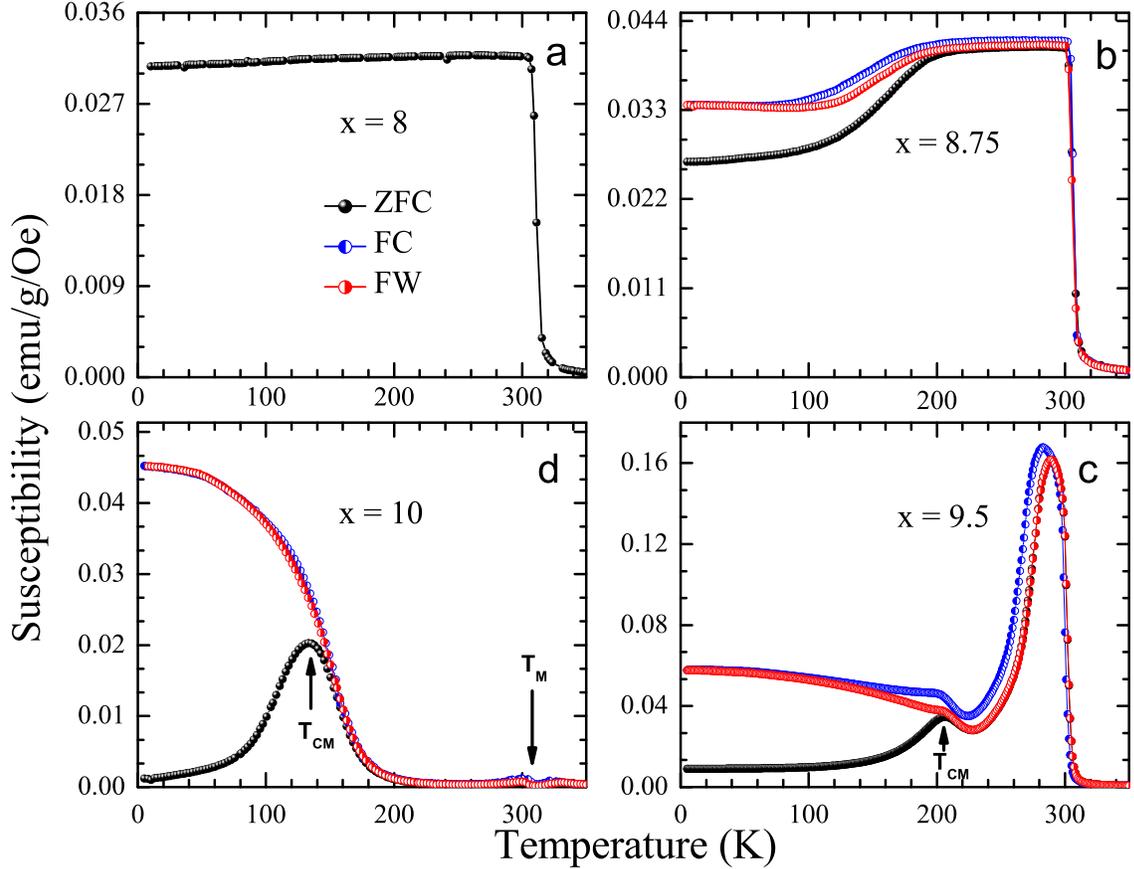}
\caption{Susceptibility as a function of temperature for the alloys Ni$_{50}$Mn$_{25+x}$Z$_{25-x}$, $x$ = 8, 8.75, $x$ = 9.5 and $x$ = 10 shown in a, b, c and d respectively during warming after cooling in zero field (ZFC), and subsequent cooling (FCC) and warming (FCW) cycles displaying martensitic transition.}
\label{fig:MT}
\end{center}
\end{figure}

Temperature dependent magnetization measurement M(T) were performed in the applied magnetic field of 50 Oe for the $x$ = 8 and $x$ = 8.75 alloys, in 100 Oe for the $x$ = 9.5 alloy and in 200 Oe for the $x$ = 10 alloy. The ferromagnetic nature of the $x$ = 8 alloy is manifested by a steep rise in its susceptibility ($\chi$ = M/H)  below its $T_C$ (see Fig. \ref{fig:MT} (a)). The $x$ = 8.75 alloy (Fig. \ref{fig:MT} (b)) also displays a ferromagnetic behavior but experiences a decrease in the susceptibility below 200 K. This decrease can be ascribed to the occurrence of a first order martensitic transition, evident from the hysteresis seen in the FCC and FCW magnetization cycles.

As a result of martensitic transition, the susceptibility of the $x$ = 9.5 alloy, in Fig. \ref{fig:MT} (c), decreases sharply and then increases to display a maximum at $T_{CM}$ = 206 K. Below this temperature, the ZFC and FC curves branch off. The ZFC data decreases with further lowering of temperature while the FC data exhibits a weak increase. Even though the martensitic transition in the $x$ = 10 alloy occurs in the paramagnetic state at $T_M$ = 333 K, its $\chi$(T) behavior in the martensitic state is similar to that of $x$ = 9.5 (Fig. \ref{fig:MT} (d)). Here, in $x$ = 10 alloy, the ZFC magnetization curve exhibits a peak at $T_{CM}$ = 135 K before approaching zero at lower temperatures. The FC curves, on the other hand, increase continuously giving the impression of the presence of a blocking temperature. It is pertinent to note that with an increase in excess Mn concentration, $T_{CM}$ decreases indicating a weakening of ferromagnetic interactions. In fact, for $x$ = 12.5 alloy $T_{CM}$ is reported to be 39 K \cite{Nevgi2019797}.

\begin{figure}[h]
\begin{center}
\includegraphics[width=\columnwidth]{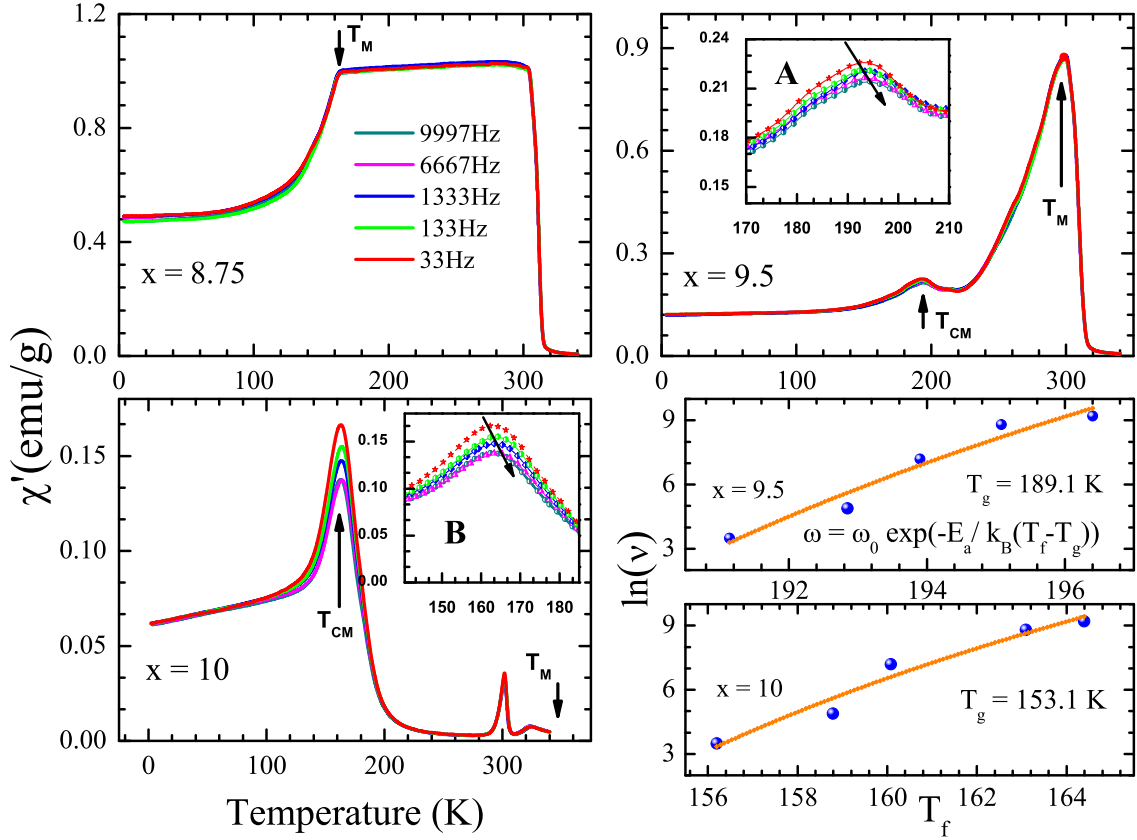}
\caption{Temperature dependent measurements of the real part of ac susceptibility for the three alloys $x$ = 8.75, 9.5 and 10 at different frequencies. Frequency dependent behaviour of T$_{CM}$ is seen in the insets A and B for the alloys $x$ = 9.5 and 10 respectively following Vogel Fulcher law.}
\label{fig:ACSUC1}
\end{center}
\end{figure}

\begin{figure}[h]
\begin{center}
\includegraphics[width=\columnwidth]{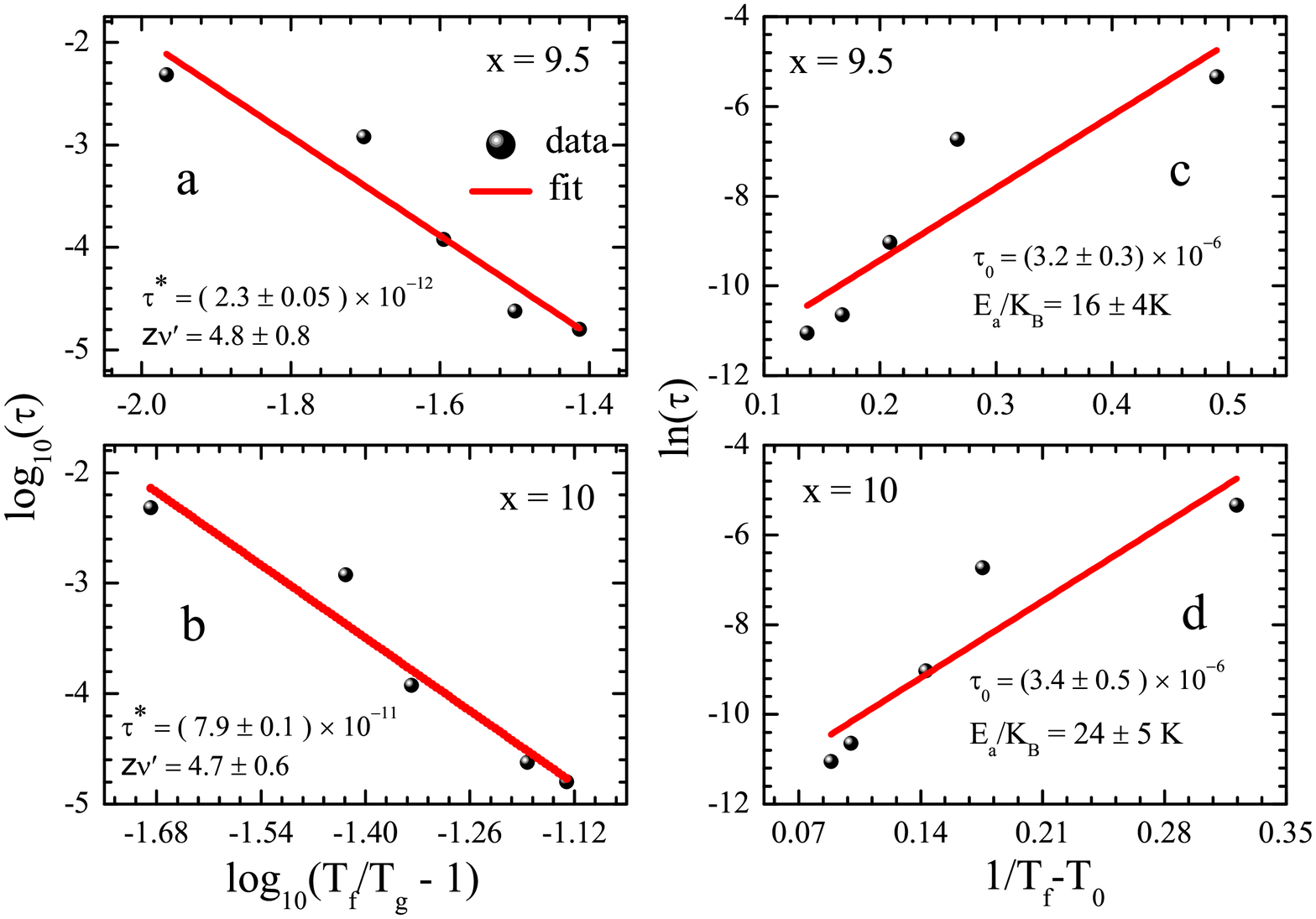}
\caption{The plots marked a and b give the best fits to the equation $\tau$ = $\tau^{*}$$({\frac{T_{f}}{T_{g}}-1})^{-z\nu^{'}}$ for the alloys $x$ = 9.5 and $x$ = 10 respectively while the Vogel Fulcher relation $\tau$ = $\tau_{0}$exp($\frac{E_{a}}{K_{B}(T_{f} - T_{0})}$) is shown in graphs marked as c and d for the two alloys.}
\label{fig:ACSUC2}
\end{center}
\end{figure}

To further probe the nature of the magnetic state in the present alloys, temperature dependent ac susceptibility measurements were performed at five different frequencies as shown in Fig. \ref{fig:ACSUC1}. The $x$ = 8.75 alloy does not display any frequency dependent behavior of the ac susceptibility signal either at its ferromagnetic transition or at the martensitic transition temperature. However, the frequency dependence of the real part of ac susceptibility, $\chi^{'}$ is clearly seen around $T_{CM}$ in $x$ = 9.5 and $x$ = 10 indicating the possibility of a non-ergodic ground state. Behavior in accordance with Vogel Fulcher law is noted for the peak temperature confirming the presence of a glassy phase in both alloys. The frequency dependence in $\chi^{'}$ is assessed by $\delta T_{f} = \frac{\Delta T_{f}}{T_{f}(\Delta\log\nu)}$ which is found to be 0.01 and 0.02 for the $x$ = 9.5 and $x$ = 10 alloys respectively. These values are larger than those expected for typical spin glass and smaller compared to those expected for a typical superparamagnet \cite{Mulder198225, Mydosh1993}. A cooperative dynamics due to inter cluster interactions are described by the Vogel Fulcher law, $\tau$ = $\tau_{0}$exp($\frac{E_{a}}{K_{B}(T_{f} - T_{0})}$) wherein $\tau_{0}$ is the time constant corresponding to characteristic attempt frequency and is related to the strength of interactions while $E_{a}$ is the activation energy of the relaxation barriers. For both the alloys, the fitting yields $\tau_{0} \sim 10^{-6}$ s and the ratio $\frac{E_{a}}{K_{B}}$ also lies between 15 to 20 as seen in Fig. \ref{fig:ACSUC2}(c) and Fig. \ref{fig:ACSUC2}(d). This is indicative of a significant coupling amongst the dynamic entities \cite{Bag4972019,Djurberg791997}. A strong inter-cluster interactions can give rise to spin-glass like cooperative freezing, and in this case, the frequency dependence of peak in  $\chi^{'}$ is expected to follow the power law divergence of the standard critical slowing down given by dynamic scaling theory, $\frac{\tau}{\tau^{*}}$ = $({\frac{T_{f}}{T_{g}}-1})^{-z\nu^{'}}$ wherein $\tau$ represents the dynamical fluctuation time scale corresponding to measurement frequency at the peak temperature of $\chi*$, $\tau^{*}$ is the spin flipping time of the relaxing entities, T$_{g}$ is the glass transition temperature in the limit of zero frequency, ${z}$ is the dynamic scaling exponent, and ${\nu^{'}}$ is the critical exponent. In the vicinity of glass transition, the spin cluster correlation length $\xi$ diverges as $\xi$ $\propto$ $({\frac{T_{f}}{T_{g}}-1})^{-\nu^{'}}$ and the dynamic scaling hypothesis relates $\tau$ to $\xi$ as $\tau$ $\sim$ ${\xi^z}$ \cite{Kumar201325,Chakrabarty201426}. The results of the best fits obtained for the $x$ = 9.5 and $x$ = 10 alloys are shown in Fig.  \ref{fig:ACSUC2}(a) and Fig.  \ref{fig:ACSUC2}(b) respectively. Here, T$_{g}$ is taken as the abscissa of the ln($\nu$)Vs T$_{f}$ plots shown in Fig.  \ref{fig:ACSUC1}. The value of $\tau^{*}$ are $2.3 \times 10^{-12}$s and $7.9 \times 10^{-11}$s and the values of  ${z\nu^{'}}$ are 4.8 and 4.7 respectively for $x$ = 9.5 and $x$ = 10 alloys. Such values have been reported in Ni-Mn alloys and are characterized as super spin glass systems \cite{Cong2512014,Liao1042014,Wang1062011}.

The frequency dependence around $T_{CM}$ in ac susceptibility following Vogel Fulcher law and scaling law advocates the presence of a glassy phase along with significant inter cluster interactions in the $x$ = 9.5 and $x$ = 10 alloys. These alloys seem to have ferromagnetic and non-ferromagnetic clusters that are actively interacting with each other leading to a glassy ground state and exhibit critical slowing down as expected from dynamical scaling theory.

\begin{figure}[h]
\begin{center}
\includegraphics[width=\columnwidth]{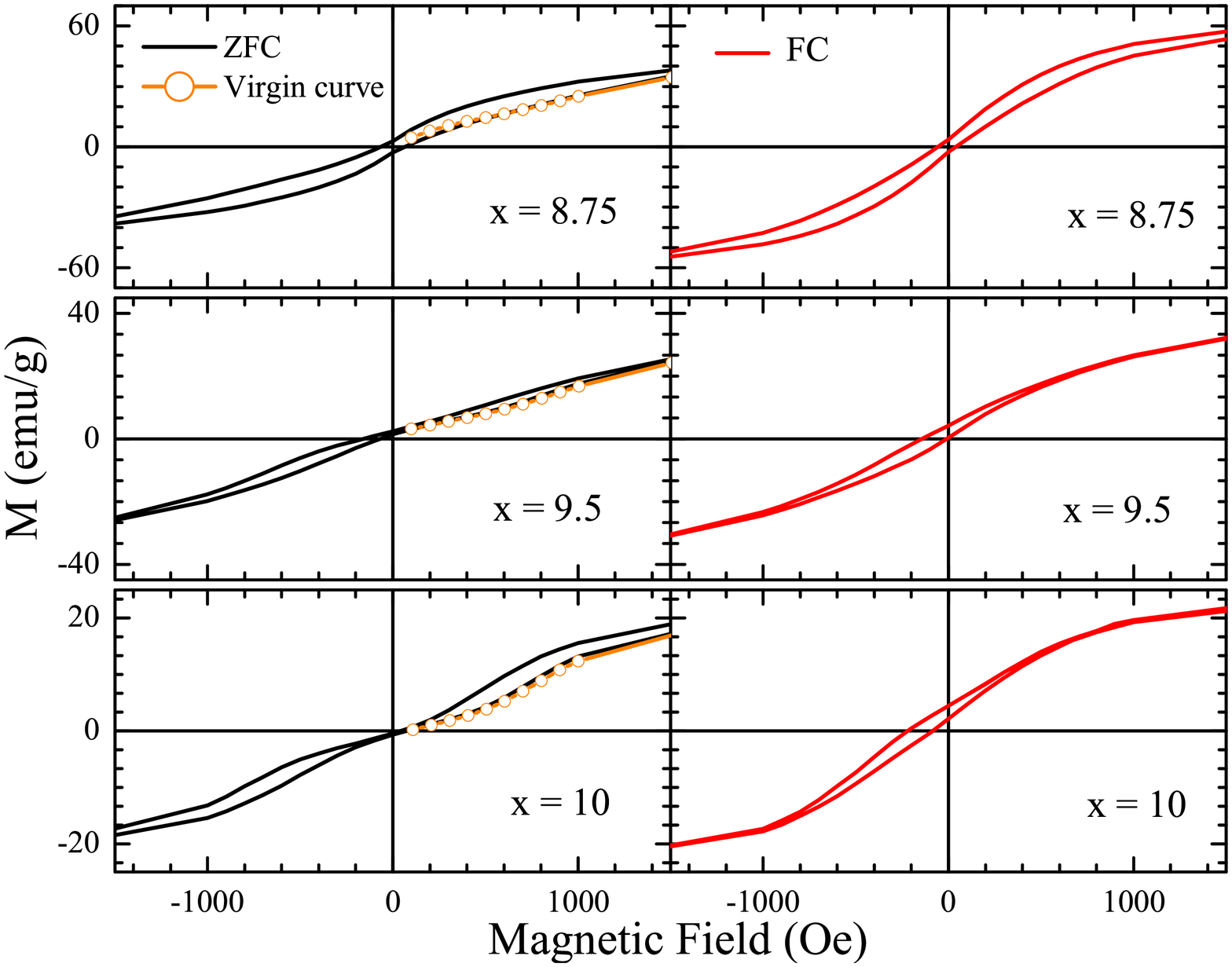}
\caption{Magnetization as a function of applied field at 5K for the three alloys $x$ = 8.75, 9.5 and 10. The dotted line in orange represents virgin curve.}
\label{fig:EBIAS}
\end{center}
\end{figure}

\begin{figure}[h]
\begin{center}
\includegraphics[width=\columnwidth]{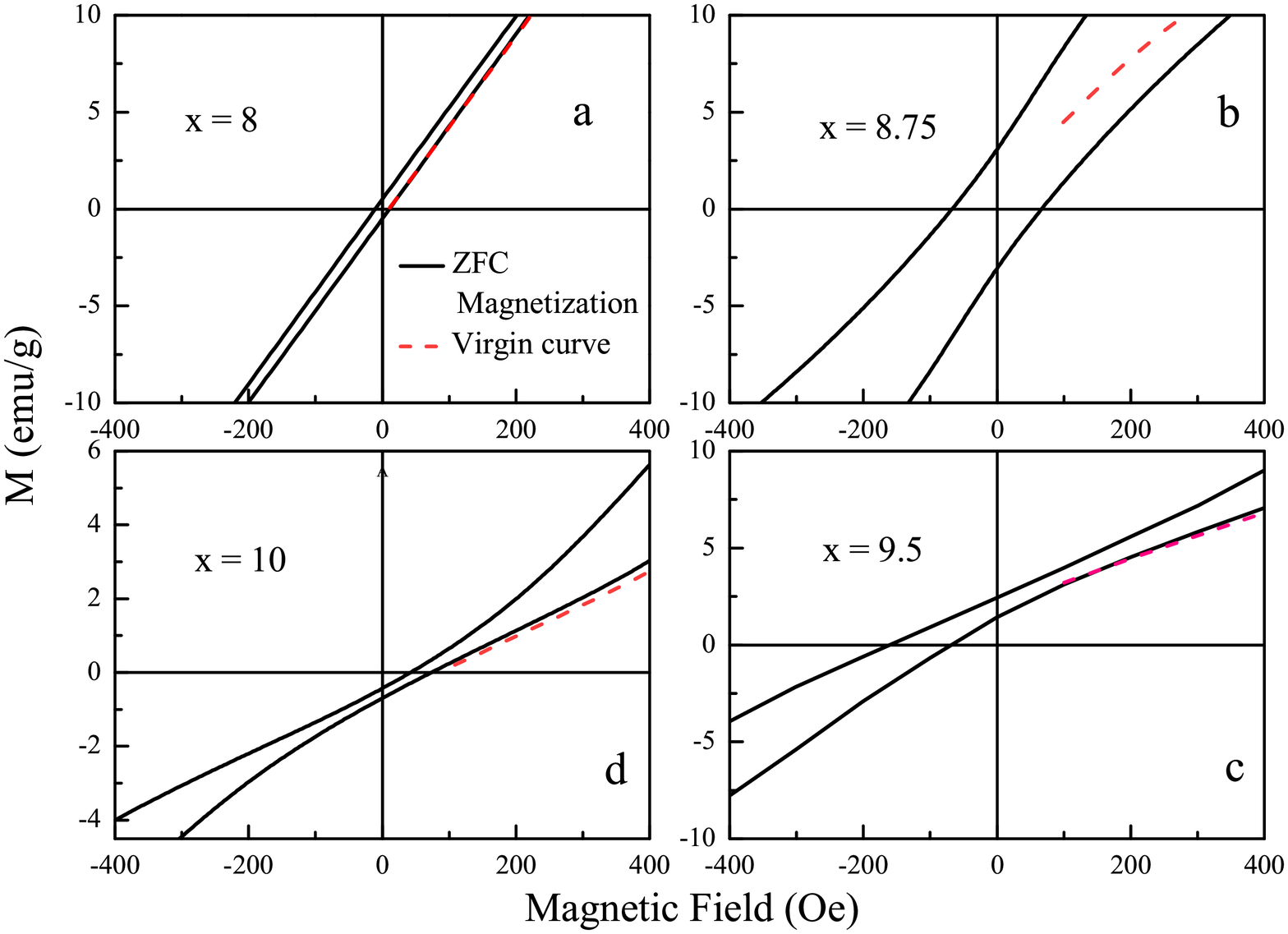}
\caption{Magnetization as a function of applied field at 5K displaying virgin curves for the four alloys $x$ = 8, 8.75, 9.5 and 10(a). The expanded view of ZFC magnetization displaying the behavior of the hysteresis loop around zero field for the alloys $x$ = 8.75, (b) 9.5, (c) and 10, (d). The inset in (c) and (d) gives a pictorial representation of magnetic ground state comprising of ferromagnetic and antiferromagnetic clusters.}
\label{fig:EBIAS1}
\end{center}
\end{figure}

A better understanding of the magnetic ground state can be obtained by studying the shape of the hysteresis loop. Fig.  \ref{fig:EBIAS} reflects the zero field cooled and field cooled M(H) data recorded at 5K for the three alloys: $x$ = 8.75, $x$ = 9.5 and $x$ = 10 while the Fig.  \ref{fig:EBIAS1} gives an expanded view of the ZFC M(H) data in the field interval of $\pm$ 400 Oe. The $x$ = 8 and 8.75 alloys display a ZFC hysteresis loop firmly around the center of the axis and the virgin curve is traced within the loop. This does not seem to be the case with the alloys $x$ = 9.5 and $x$ = 10. (Fig. \ref{fig:EBIAS1}) In both these alloys, the loop appears to be displaced vertically up and down respectively with the virgin curve lying outside the loop. These are ascribed as signatures of the presence of ferromagnetic and antiferromagnetic interacting clusters \cite{Bhobe200841}. In case of $x$ = 9.5, it appears that the coupling between the ferromagnetic and antiferromagnetic clusters is favorable along the +H direction due to which the magnetization assumes a higher magnitude in that direction and the loop appears to be shifted up while in case of $x$ = 10 the situation appears to be reversed causing the loop to shift down. The presence of exchange bias in M(H) loops also assures the presence of both ferromagnetic and antiferromagnetic interactions and the value of exchange bias field increases from -78.96 Oe in $x$ = 9.5 to -149.23 Oe in $x$ = 10. The increasing values of exchange bias field with increasing Mn at In site ($x$) perhaps hints at the growth of antiferromagnetic clusters in these alloys.

To correlate the existence of magnetic glass with a possible presence of strain glass, the alloys with  $x$ = 8.75 and $x$ = 9.5 were further investigated for frequency dependent elastic properties by performing the dynamical mechanical analyzer (DMA) studies in the temperature range of 100 K to 400 K (not shown). No frequency dependence of storage modulus or loss were visible over a frequency range 0.1 Hz to 10 Hz ruling out the possibility of existence of the strain glass in these alloys.

\begin{figure}[h]
\begin{center}
\includegraphics[width=\columnwidth]{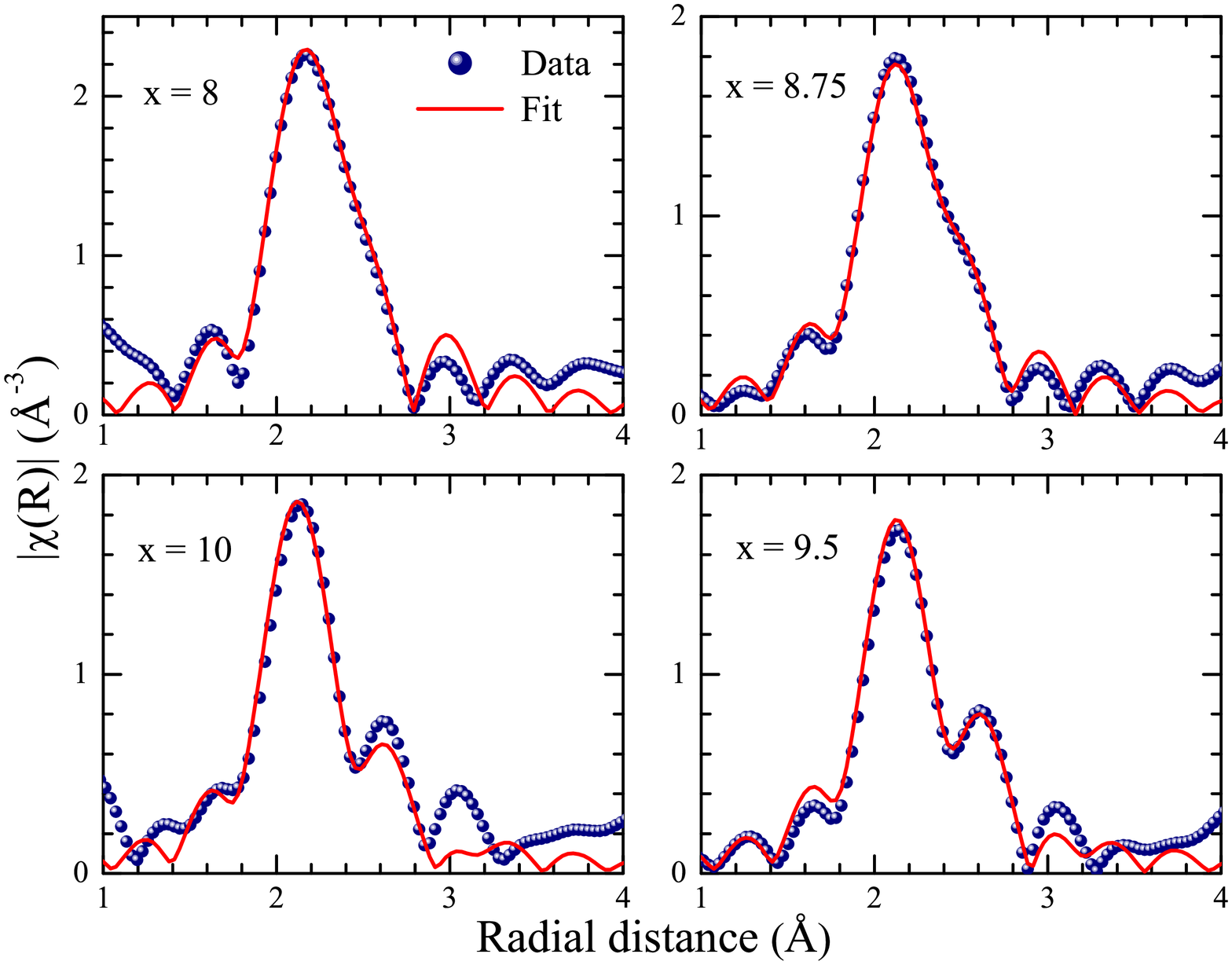}
\caption{Magnitude of the Fourier transformed EXAFS spectra obtained at Ni K edge in the $x$ = 8, 8.75, 9.5 and 10 alloys at 50K}
\label{fig:EXAFS2}
\end{center}
\end{figure}

\begin{figure}[h]
\begin{center}
\includegraphics[width=\columnwidth]{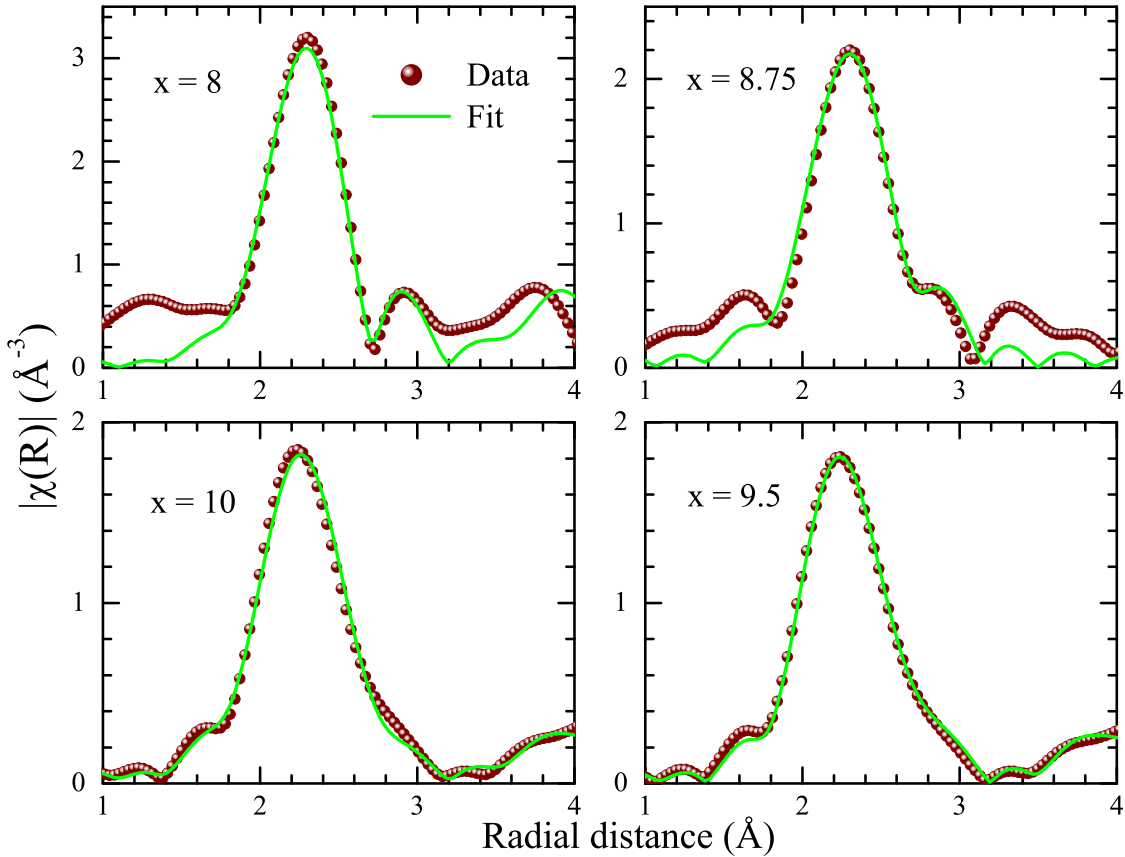}
\caption{Magnitude of the Fourier transform of the Mn K edge EXAFS spectra in the alloys with $x$ = 8, 8.75, 9.5 and 10 at 50K}
\label{fig:EXAFS3}
\end{center}
\end{figure}

\begin{figure}[h]
\begin{center}
\includegraphics[width=\columnwidth]{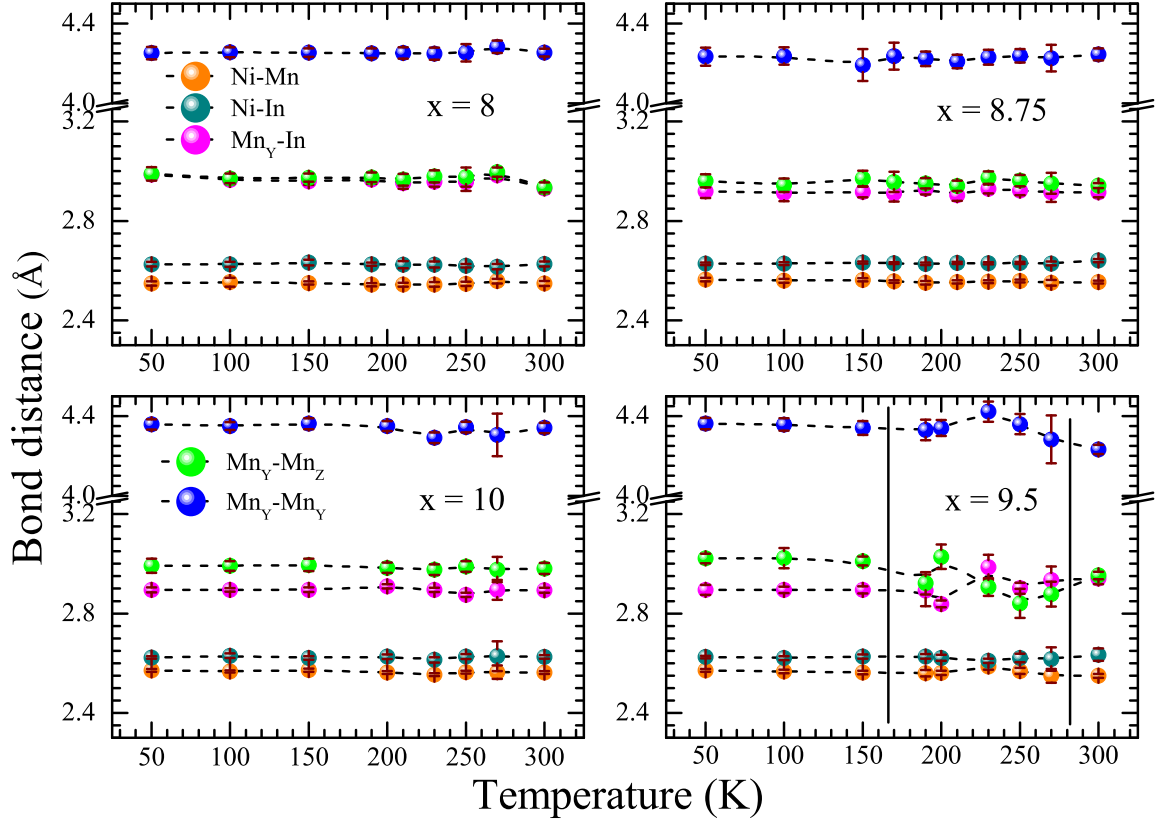}
\caption{Variation of bond distances with temperature for the alloy compositions $x$ = 8, 8.75, 9.5 and 10. The oscillations in bond distances seen in the region between the two dashed lines highlights the martensitic transformation region in $x$ = 9.5.}
\label{fig:EXAFS1}
\end{center}
\end{figure}

\begin{figure}[h]
\begin{center}
\includegraphics[width=\columnwidth]{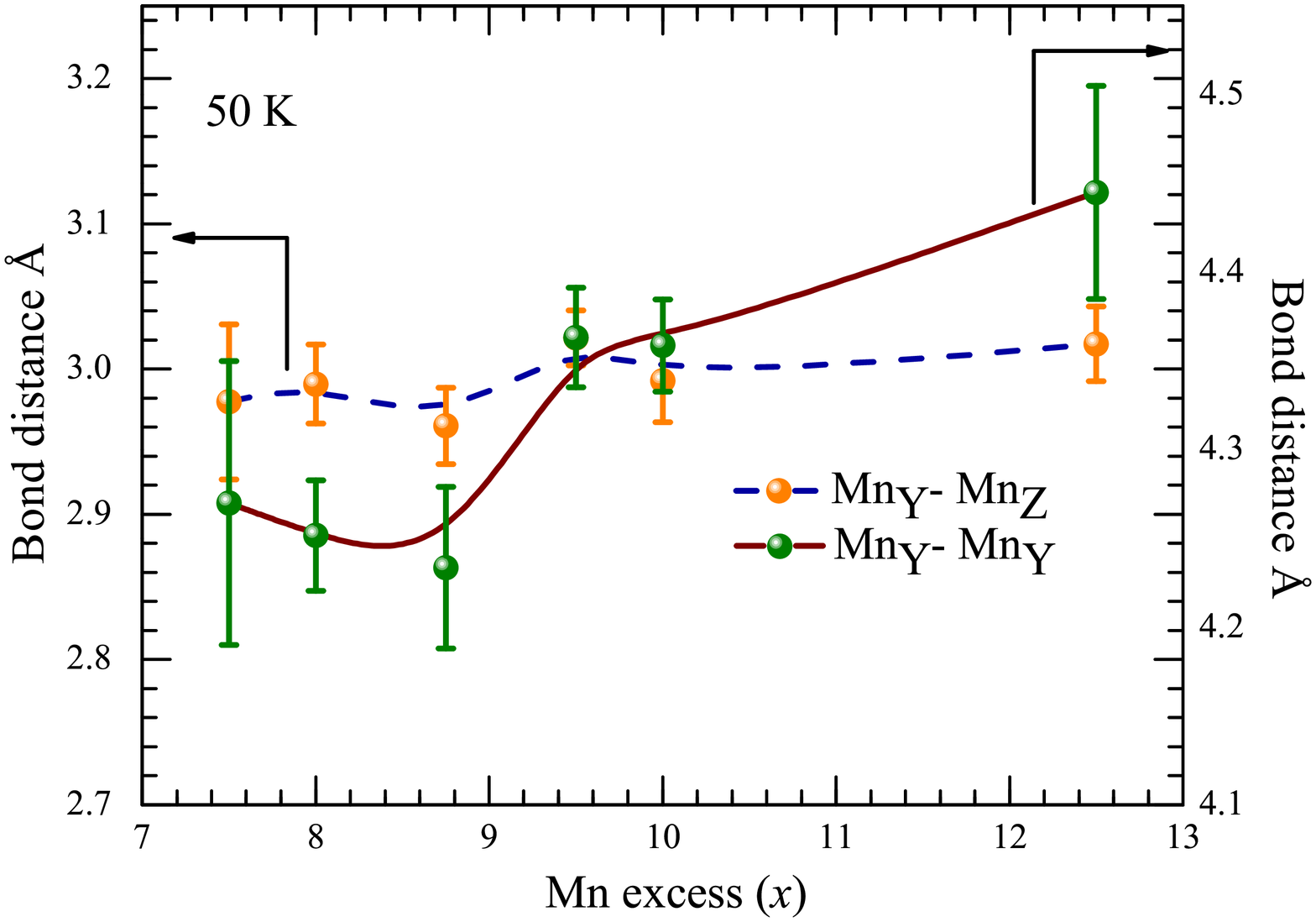}
\caption{Variation of Mn-Mn bond distances with Mn excess concentration $x$ at 50 K}
\label{fig:EXAFS4}
\end{center}
\end{figure}

In the Ni$_{50}$Mn$_{25+x}$In$_{25-x}$ alloys, the ferromagnetic correlations are mediated via RKKY interaction between the Mn atoms in its own sublattice \cite{Buchelnikov782008} while the antiferromagnetic correlations arise due to an exchange type interaction between Mn$_Y$ (Mn in its sublattice) and Mn$_Z$ (Mn at In site) atoms mediated via Ni atom \cite{Priolkar872013}. Therefore, a complete understanding of magneto-structural properties demands a careful study of the local structure owing to the fact that Mn excess alloys exhibit local structural distortions even in the austenitic phase \cite{Lobo201096}. As a result, EXAFS data analysis was performed for all the four alloy compositions at both Ni K and Mn K edges in the temperature range of 50 K to 300 K. EXAFS data at Ni and Mn K edges were analyzed together using a common structural model. The analysis was carried out using 14 independent parameters. The amplitude reduction factor ($S_0^2$) for the two data sets were obtained from the analysis of respective metal foils and were kept fixed during the analysis. The magnitude of Fourier transform (FT) of Ni K edge EXAFS spectra shown in Fig. \ref{fig:EXAFS2} include contribution from the nearest neighbors Mn and In atoms at $\sim$ 2.5\AA~ with their coordination number fixed as per the composition ratio and from the next nearest neighbor Ni atoms ($\sim$ 3.0\AA). Mn K  edge EXAFS spectra (Fig. \ref{fig:EXAFS3}), on the other hand, are fitted considering the nearest neighbor, Ni ($\sim$ 2.5\AA), the next nearest neighbor, In and Mn$_Z$ ($\sim$ 3.0\AA) and the third neighbor, Mn$_Y$ at ($\sim$ 4.2\AA) as backscattered atoms. Best fits at all temperatures were obtained only after relaxing the constraints imposed by the cubic austenitic structure as described earlier \cite{Bhobe200674}. The variation of bond distances with temperature represented in Fig. \ref{fig:EXAFS1} clearly shows that the nearest neighbor Ni--Mn distance is shorter than the  Ni--In bond distance irrespective of excess Mn content. In the case of alloys $x$ = 8 and $x$ = 8.75, the bond distances of Mn$_Y$--In and Mn$_Y$--Mn$_Z$ are almost equal and in accordance with their cubic crystal structure. Here Mn$_Y$ represents Mn in its sublattice while Mn$_Z$ represents Mn in Z (In) sublattice of X$_2$YZ Heusler structure. The nearly equal Mn$_Y$--In and Mn$_Y$--Mn$_Z$ distances and shorter Ni--Mn distance imply that the structural distortions due to replacement of In atoms by relatively smaller Mn atoms are restricted only to nearest neighbor correlations. As the Mn concentration is increased beyond $x$ = 8.75, the Mn$_Y$--Mn$_Z$ distance exceeds  Mn$_Y$--In bond distance especially in the martensitic phase indicating release of structural strain and lowering of crystal symmetry. A comparison of 50 K values of the third neighbor Mn$_Y$--Mn$_Y$ distance with the Mn$_Y$--Mn$_Z$ distance for a wider range of excess Mn concentration, $7.5 \le x \le 12.5$, presented in Fig. \ref{fig:EXAFS4} shows that the  Mn$_Y$--Mn$_Y$ distance increases rapidly in the martensitically transforming alloys while the Mn$_Y$--Mn$_Z$ bond distance remains nearly constant throughout the concentration range.

\section{Discussion}

The above studies depict that an increase in excess Mn concentration in Ni$_{50}$Mn$_{25+x}$In$_{25-x}$ alloys lead to the occurrence of the martensitic transformation at a critical value of $x$ = 8.75. Though the austenitic structure is retained down to the lowest measured temperature, alloys with slightly higher Mn concentration ($x \geq 9.5$) transform completely to an incommensurate 7M monoclinic martensitic structure. There are no signatures  of any pre-transformation phases, like strain glass, in any alloy compositions with either $x \geq 8.75$ or $x < 8.75$.

Concomitant with this structural transformation, magnetic properties also change drastically. The martensitic alloy compositions ($x \ge 9.5$), whether transforming in ferromagnetic or paramagnetic state, share some common features. In the martensitic state, the magnetization increases with ZFC and FC curves exhibiting irreversible behavior below a characteristic temperature $T_{CM}$. With an increase in  Mn content, $T_{CM}$ decreases, the ground state transforms from an ordered ferromagnetic state to a state with ferromagnetic and antiferromagnetic spin clusters with glassy dynamics. The alloys exhibit an exchange bias with the exchange bias field increasing with increasing Mn content. The decrease in $T_{CM}$ and increasing exchange bias field suggests the growth of antiferromagnetic clusters at the expense of ferromagnetic clusters. Neutron diffraction studies on Co doped Ni-Mn-Ga alloys have shown that antiferromagnetic interactions in such Heusler alloys are not due to Mn$_Y$--Mn$_Z$ interactions but arise from an antiferromagnetic moment on the Ni atoms at the X site of the X$_2$YZ Heusler structure \cite{Orlandi942016}. The role of Ni in the antiferromagnetic interactions in Mn rich Heusler alloys was also highlighted earlier from XMCD studies \cite{Priolkar872013}. Invariant Mn$_Y$--Mn$_Z$ bond distance and an increasing Mn$_Y$--Mn$_Y$ bond distance with increasing Mn concentration seen from the EXAFS analysis supports the view that Mn$_Y$--Mn$_Z$ interactions alone are not responsible for antiferromagnetic interactions. Such a role of Ni atoms in the antiferromagnetic interactions coupled with the observed Ni--Mn bond distance to be shorter than Ni--In bond distance points towards a possibility of formation of two structural variants at the local level, the ferromagnetic Ni$_{50}$Mn$_{25}$In$_{25}$ and the antiferromagnetic Ni$_{50}$Mn$_{50}$. Temper annealing of Ni$_{50}$Mn$_{25+x}$In$_{25-x}$ alloys have shown the disintegration of the modulated martensitic structure into Heusler L2$_1$ and tetragonal L1$_0$ phases \cite{Cakir1272017,Dincklage82018}. The local segregation of different structural variants could be responsible for the observed ferromagnetic and antiferromagnetic clusters leading to a super spin glassy ground state.

The absence of any non-ergodic pre-transition phases like strain glass despite the presence of magnetic clusters can be related to the ability of the Heusler structure to accommodate the lattice strain caused by the difference in sizes of Mn and In atoms occupying the Z site of Heusler structure. The cubic symmetry of the austenitic Heusler structure demands Ni--Mn and Ni--In bond distances to be equal. But the analysis of EXAFS data indicates Ni--Mn bond distance to be shorter than the Ni--In bond distance in all non stoichiometric, Ni$_{50}$Mn$_{25+x}$In$_{25-x}$ alloys. The nearly equal next nearest neighbor Mn$_Y$--In and Mn$_Y$--Mn$_Z$ bond distances help in preserving the cubic austenitic order. Such a structural distortion builds up a lattice strain which is relieved by a transformation to the martensitic state beyond the critical concentration of Mn replacing In at the Z sites. At this point, a difference between Mn$_Y$--Mn$_Z$ and Mn$_Y$--In bond distances are also seen. The effect of strain accommodation up to the critical concentration is also reflected in the behavior of ferromagnetic Mn$_Y$--Mn$_Y$ distance in Fig.\ref{fig:EXAFS4}. The Mn$_Y$--Mn$_Y$ distance exhibits a relatively rapid increase beyond the critical concentration ($x$ = 8.75). The similarity between the variation of Mn$_Y$--Mn$_Y$ distance and the $T_M$ as a function of $x$ suggests a strong connection between the magnetic and structural degrees of freedom. It appears that the ability of the Heusler structure to adapt to the strain caused by the size difference between In and doped Mn atoms is due to the presence of ferromagnetic interactions between Mn atoms and is perhaps the reason for the absence of any non ergodic structural pre-transition phases in Ni$_{50}$Mn$_{25+x}$In$_{25-x}$ alloys.

\section{Conclusions}

In conclusion, the cubic Heusler structure of Ni$_{50}$Mn$_{25+x}$In$_{25-x}$ has the ability to accommodate the lattice strain caused by the replacement of larger In atom with smaller Mn atom up to a critical concentration. Beyond the critical value of $x$, the cubic structure relieves the strain by undergoing a martensitic transition. Along with this structural relaxation process, the magnetic ground state transits from a ferromagnetically ordered to a super spin glass state with antiferromagnetic and ferromagnetic clusters.  Even though, with increasing Mn doping, the magnetic transition is non-ergodic, the ferroelastic transition is ergodic. This appears to be due to the ability of the Heusler structure to accommodate strain and retain ferromagnetism in Ni$_{50}$Mn$_{25+x}$In$_{25-x}$.

\section*{Acknowledgements}
KRP and RN would like to thank the Science and Engineering Research Board, Govt. of India under the project SB/S2/CMP-0096/2013 for financial assistance, Department of Science and Technology, Govt. of  India for the travel support within the framework of India\@ DESY collaboration. RN acknowledges the Council of Scientific and Industrial Research, Govt. of India for Senior Research fellowship and acknowledges MAECI fellowship from the Govt. of Italy for work done at Parma. Edmund Welter and Ruidy Nemausat are thanked for experimental assistance at P65 beamline, PETRA III, DESY Hamburg. The authors also thank Fabio Orlandi (STFC,RAL,ISIS) for a fruitful discussion on the manuscript. 

\bibliographystyle{iopart-num}
\bibliography{Ref}

\end{document}